\documentclass{andp2012}
\usepackage[english]{babel}
\setcopyrightyear{2015}%
\Volume{XXX}%
\Year{2015}%
\category{Original Paper}
\title{The Higgs boson, Supersymmetry and Dark Matter}
\subtitle{Relations and Perspectives}
\author[A. Arbey]{Alexandre Arbey\inst{1}}
\author[M. Battaglia]{Marco Battaglia\inst{2}
\footnote{Corresponding author\quad E-mail:~\textsf{marco.battaglia@ucsc.edu}}}
\author[F. Mahmoudi]{Farvah Mahmoudi\inst{1}}
\address[1]{Universit{\' e} de Lyon, Universit{\' e} Lyon 1;
  Centre de Recherche Astrophysique de Lyon, Saint-Genis Laval Cedex;
  Ecole Normale Sup{\' e}rieure de Lyon, France and
  CERN, CH-1211 Geneva 23, Switzerland}
\address[2]{Santa Cruz Institute of Particle Physics, University of California,
  Santa Cruz, CA 95064, USA and
  CERN, CH-1211 Geneva 23, Switzerland}
\shortauthors{A. Arbey, M. Battaglia, F. Mahmoudi}
\begin{abstract}
  The discovery of a light Higgs boson at the LHC opens a broad program
  of studies and measurements to understand the role of this particle in
  connection with New Physics and Cosmology. Supersymmetry is the best
  motivated and most thoroughly formulated and investigated model of New
  Physics which predicts a light Higgs boson and can explain dark matter.
  This paper discusses how the study of the Higgs boson connects with
  the search for supersymmetry and for dark matter at the LHC and at a
  future $e^+e^-$ collider and with dedicated underground dark matter
  experiments.  
\end{abstract}
\shortabstract
\begin{document}
\maketitle

\section{Introduction}

The discovery of a light Higgs boson~\cite{ATLAS:2012zz,CMS:2012zz} has opened a new chapter
in the search for new physics and its connection to cosmology through dark matter (DM).
The properties of the Higgs boson may be sensitive to new physics beyond the Standard Model
(BSM), either because the particle observed by ATLAS and CMS is part of an extended Higgs
sector or because new particles may modify its couplings and decay rates compared to those
predicted by the Standard Model (SM). If dark matter is due to a weakly-interacting massive
particle (WIMP), the Higgs boson most likely couples to it. In this case it may have a major
role in mediating the WIMP interactions, which have set its relic density in the universe to
the observed level, and those which are being exploited to detect dark matter interactions
in underground experiments. In all these cases, the study of the Higgs boson properties,
the search for BSM physics at colliders and the direct searches for dark matter at dedicated
experiments will likely shape a new picture of particle physics and cosmology, which integrates
the Higgs boson as a central piece.

Supersymmetry (SUSY) has been so far the best motivated and most thoroughly formulated and
investigated model of BSM physics, which predicts the Higgs boson to be light and naturally
incorporates dark matter.
Originally introduced to solve the hierarchy problem of the SM, SUSY with conserved R-parity
offers a remarkable signature of its production in collider experiments: SM particles + missing
energy. Its lightest particle, which is stable, takes the role of the WIMP, explaining the
observed DM relic density for a broad range of SUSY model parameters. Even if SUSY is not
realised in nature, its study is of great interest since it represents an attractive template
of BSM theories with a conserved quantum number.

In this paper, we discuss the current status and future perspectives for the study of the Higgs
boson in relation to BSM physics, here taken to be supersymmetric, and dark matter.
The results of the Higgs property measurements and of the searches for BSM physics conducted
on the data collected by the LHC experiments at 7 and 8~TeV and the bounds set on the WIMP scattering
cross section by the Xenon~\cite{Aprile:2012nq} and LUX~\cite{Akerib:2013tjd} experiments already
provide a formidable set of measurements and constraints.
The forthcoming LHC run at 13-14~TeV will extend the number of processes tested and the accuracy of
their measurements in the study of the Higgs boson profile~\cite{Dawson:2013bba}. The sensitivity to
new particles will be greatly boosted by the increase in both energy and integrated luminosity,
significantly extending it well beyond 1~TeV for the majority of the SUSY states.
The anticipated sensitivity of the LZ data will push the WIMP scattering cross section bounds
down by more than two orders of magnitude~\cite{Malling:2011va,Cushman:2013zza}.

The past two decades have been rich in technical and conceptual developments, which now make
it possible to envisage new colliders of unprecedented luminosity and
energy~\cite{Barletta:2014vea,Barletta:2014nka}. An $e^+e^-$ collider of luminosity in excess to
10$^{34}$ cm$^{-2}$ s$^{-1}$ and energy up to $\sim$1~TeV based on superconducting cavities is
technically feasible and the ILC project is now being considered for construction~\cite{Behnke:2013xla}.
The luminosity of the LHC can be increased by a factor of$\sim$10 to deliver an astonishing 3~ab$^{-1}$
of integrated luminosity with the HL-LHC program by $\sim$2035. A circular hadron collider with collision
energy in the range 80-100~TeV can be realistically envisaged for the 2040s (FCC-hh), housed in a tunnel
which could also be used for an $e^+e^-$ collider of energy up to the $HZ$ threshold
(FCC-ee)~\cite{Zimmermann:2014qxa}.
On the time scale of the HL-LHC and a possible ILC, DM direct detection experiments of third generation
could explore the region of scattering cross section values down to the limit where neutrino
scattering becomes an irreducible background~\cite{Cushman:2013zza}.

Here, we do not attempt to be exaustive in discussing the complex inter-relations between the
Higgs sector, SUSY and DM. Rather, we wish to highlight some aspects, emerged from our studies
of parameteric scans of the SUSY phase space which can be investigated in some details with the data
already at hand and offer promising perspectives for the forthcoming LHC runs as well as for future
colliders.  First, we want to highlight that the SUSY squark and gluino contributions to the light Higgs
decay branching fractions may remain sizeable for SUSY particle masses well beyond the
sensitivity of the LHC direct searches. For appropriate combinations of the SUSY parameters, the
precision study of the Higgs decay properties at the LHC and an $e^+e^-$ collider may thus reveal
SUSY signals, even if the SUSY mass scale is beyond the LHC kinematic reach. Then, we
consider the interplay between the Higgs sector and dark matter, if this is of supersymmetric nature.
The contribution of the Higgs to the neutralino scattering cross section and the complementarity of 
the region of parameter space, where the Higgs couplings and the WIMP neutralino scattering can give
signals from BSM physics, offer new opportunities to test SUSY using data from collider and dark
matter direct detection experiments.


\section{The phenomenological MSSM, dark matter and colliders}

The minimal supersymmetric extension of the SM (MSSM) is the most economical implementation of
supersymmetry, which introduces only two Higgs doublets resulting in five physical Higgs particles.
Additional Higgs singlets or doublets can also be considered, for example in the so-called NMSSM, and
result in an increase of the complexity of the relations between the Higgs, supersymmetric and SM sectors
with the increased number of parameters of the model. However, these extensions are not required,
nor suggested, by the present data and here we consider the minimal SUSY implementation in the MSSM.

The study of the relations between the Higgs sector, new physics and dark matter in the MSSM can
be aptly performed in the so-called phenomenological MSSM (pMSSM).
This 19-parameter implementation of the MSSM offers the freedom and generality of a model where all
the SUSY particle masses are free and independent, while keeping the number of parameters manageable
for extensive scans~\cite{Djouadi:1998di}. The pMSSM is becoming a widely recognised framework to
evaluate the impact of the LHC bounds on the MSSM viability and is being used in both phenomenological
studies~\cite{AbdusSalam:2009qd,Sekmen:2011cz,Arbey:2011un,Arbey:2012dq,CahillRowley:2012kx,Arbey:2012bp,
  Cahill-Rowley:2014boa}
and by the LHC experiments~\cite{CMS:2013rda,CMS:2014mia}.
In this study, we take the lightest neutralino to be the LSP and the values of all SUSY particle
masses are varied, independently, up to 5~TeV, the SUSY trilinear couplings in the range -15 to
15~TeV and 2~$< \tan \beta <$~60 in flat scans. These ranges are important because the values of
the fractions of pMSSM points allowed or excluded by the various constraints discussed later in the
paper depend on them.
Details on the programs used for performing the pMSSM scans, computing the SUSY spectra and the related
observables and the constraints imposed on the accepted pMSSM points from low energy data
and flavour physics can be found in \cite{Arbey:2011un}. Only points with the lightest Higgs mass in the
range 123 $< M_h <$ 128~GeV, compatible with the ATLAS and CMS measurements when accounting for systematic
and model uncertainties, are accepted.

\begin{figure}[ht!]
  \begin{center}
    \vspace*{-0.5cm}
     \begin{tabular}{c}
       \includegraphics[width=0.70\columnwidth]{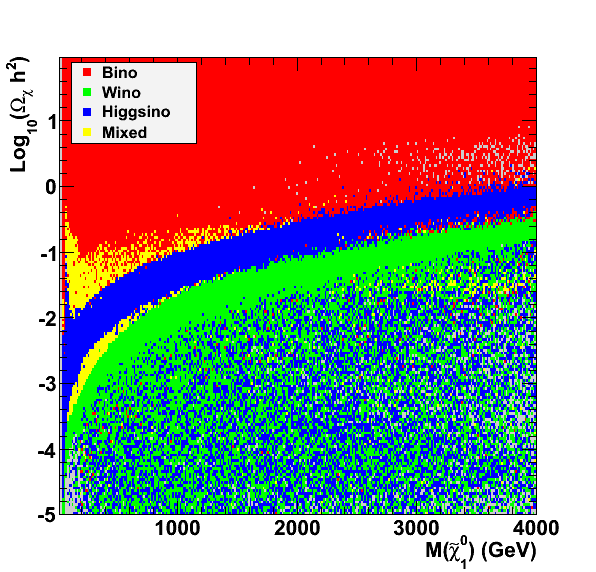} \\
           \vspace*{-0.5cm} \\
       \includegraphics[width=0.70\columnwidth]{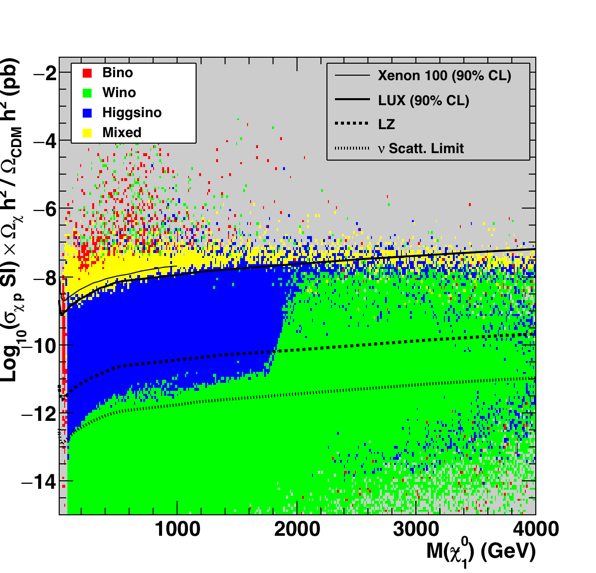} \\
     \end{tabular}
     \vspace*{-1.25cm}
   \end{center}
   \caption{   \label{fig:DMMN1} \col Neutralino relic density (upper panel) and scaled scattering
     cross section (lower panel) in the MSSM as a function of its mass. The colours indicate
     the nature of the neutralino LSP.}
   \vspace*{-0.5cm}
\end{figure}

The pMSSM gives a clear view onto the scenarios of neutralino dark matter without the biases
introduced by the highly constrained versions of the MSSM, such as the CMSSM, which were widely used
for benchmark studies and the studies of supersymmetric dark matter before the start of the LHC.
The WIMP ``miracle'' paradigm has that a particle of mass $\sim$100~GeV with typical weak-interaction
couplings generates exactly the correct amount of dark matter relic density observed in the universe.
In reality, within a well-defined model the couplings are controlled by several model parameters and
interactions with other new particles may alter the WIMP density, resulting in significant modifications
of the range of the viable WIMP masses.
The annihilation cross section varies with the nature of the neutralino LSP and co-annihilation may
reduce the neutralino density, we observe that $\chi^0_1$ masses as low as 10~GeV (for bino-like neutralinos)
and as large as 3.5~TeV (for wino-like neutralinos) can be comfortably accomodated by the cosmic microwave
background (CMB) data, as illustrated in Figure~\ref{fig:DMMN1} obtained from the results of our pMSSM scans.
We test two sets of $\Omega_{\chi} h^2$ constraints for the pMSSM points with neutralino LSP.
First, we apply a strict bound requiring the neutralino relic  density is in agreement with the dark matter
relic density obtained from the PLANCK data, $\Omega_{CDM} h^2$~\cite{Ade:2013zuv}, allowing for systematic
uncertainties, i.e.\ $0.090 < \Omega_{\chi} h^2 < 0.163$. Then, we also consider a looser constraint, by
requesting that the neutralino relic density does not exceed the upper bound on the dark matter density from
the CMB data, i.e. $10^{-5} < \Omega_{\chi} h^2 < 0.163$,
again after acconting for systematic uncertainties. This allows for additional sources of dark matter,
other than the neutralino LSP. In this case, it is appropriate to rescale the $\chi$ scattering cross section
for the neutralino LSP case by the ratio of the neutralino relic density to the CMB value, referred to in the
following as ``scaled scattering cross section''.

For this study we consider the constraints derived from the LHC data obtained in the 7+8~TeV runs,
the results from direct DM searches at LUX\cite{Akerib:2013tjd} and the perspectives for the LHC Run~2,
the HL-LHC program, the ILC and LZ. The expected accuracies for the determination of the Higgs properties
for the future programs are taken from the compilation in~\cite{Dawson:2013bba}.
The ATLAS and CMS experiments at the LHC have pursued a vast program of searches for SUSY particles.
We test the compatibility of the accepted pMSSM points with the bounds implied by a selection of these
searches. Event samples are generated and a parametric simulation for the event reconstruction
is performed. We apply the signal selection criteria for the ATLAS analyses in the
jets+MET~\cite{ATLAS-CONF-2013-047}, $b$-jets+MET~\cite{ATLAS-CONF-2013-053},
$\ell$(s)+($b$)jets+MET\cite{Aad:2014qaa,ATLAS-CONF-2013-037},
2 \& 3 $\ell$s+MET~\cite{ATLAS-CONF-2013-049,ATLAS-CONF-2013-035},
$\ell$+$bb$+MET~\cite{ATLAS-CONF-2013-093} channel, the ATLAS and CMS searches in the mono-jet
channel~\cite{CMS-EXO-12-048,ATLAS-CONF-2013-068} and the CMS search for
$H/A \rightarrow \tau \tau$~\cite{CMS-HIG-13-021}. The number of SM background events in the
signal regions are taken from the estimates by the experiments and rescaled accordingly for
the future projections. The 95\% confidence level (C.L.) exclusion of each SUSY point in
presence of background only is determined using the CLs method~\cite{Read:2002hq}. 

\section{Higgs sensitivity to SUSY}

The lightest MSSM Higgs boson, $h^0$, represents the supersymmetric counterpart of the SM Higgs boson,
$H^0_{SM}$. It is well known that the effects of the extended Higgs sector and loops of SUSY particles,
mostly $\tilde{t}$, $\tilde{b}$, $\tilde{\tau}$ and $\chi^{\pm}$ may result in shifts of the
$h^0$ couplings to fermions and gauge bosons compared to those of the SM Higgs boson and thus affect
its decay widths and branching fractions~\cite{Djouadi:2005gj,Arbey:2012bp}. These effects, if
detected in the precision study of the Higgs profile at the LHC and an $e^+e^-$ collider, may not
only indirectly signal the existence of supersymmetry but also point to the value of
some of the SUSY parameters.

In our study we test the compatibility of the Higgs properties for each accepted pMSSM point
with those predicted for the SM Higgs by computing the $\chi^2$ probability
for the Higgs signal strengths normalised to its SM expectation,
$\mu = (\sigma \times {\mathrm{BR}}) / (\sigma \times {\mathrm{BR}}) |_{SM}$,
of the $h \rightarrow bb$, $\tau \tau$, $WW$, $ZZ$ and $\gamma \gamma$ channels for the LHC and for the
branching fractions of the $h \rightarrow bb$, $cc$, $\tau \tau$, $WW$, $ZZ$ and $\gamma \gamma$ channels
for an $e^+e^-$ collider.

An interesting question arises concerning the indirect sensitivity to new physics through the study
of the Higgs branching fractions compared to the direct sensitivity from LHC searches in the MET and
other channels. The rapid increase of the mass bounds for SUSY particles at the LHC and the expected
$1/M^2$ decrease of the effect of new particles to the Higgs couplings brings into question the role
of the precision study of the Higgs profile by the time when hundreds of fb$^{-1}$ of 14~TeV will
have been collected by ATLAS and CMS. The answer to this question depends largely on the particle
under consideration.

The extended Higgs sector of the MSSM modifies the lightest Higgs couplings to up- and
down-type quarks by terms which scale inversely with the CP-odd $A$ boson mass as
$2 M_Z^2 / M_A^2 \tan^2 \beta$ and $2 M_Z^2 / M_A^2$, respectively~\cite{Djouadi:2005gj}.
These give an indirect sensitivity to the scale of $M_A$, if deviations in the branching fractions to
up- and down-type quarks are detected, or a lower bound on $M_A$, if the coupling properties agree
with the SM predictions. The size of the effect on the coupling decreases as $1/M_A^2$.
The direct sensitivity to the $A^0$ (and $H^0$) boson at the LHC comes,
at present, mostly from the $pp \rightarrow A \rightarrow \tau^+ \tau^-$ process.
The $bbH$ associate production and gluon fusion processes~\cite{Muhlleitner:2010zz} result in a decrease
of the total cross section $\propto \tan \beta$ up to the point where the $b$ loops take over and the
cross section increases. The decay branching fraction is $\propto \tan \beta$ for $\tan \beta < 10$.
All this makes the bounds from the $\tau \tau$ final state particularly strong at large values of
$\tan \beta$ but quite unconstraining at $\tan \beta \simeq$ 10.

Looking to the modifications of the Higgs couplings to fermions induced by loops of
strongly-interacting SUSY particles, the case of the $\Delta_b$-induced shift of the Higgs coupling
to $bb$, and thus all the Higgs branching fractions, is of special importance. The SUSY contribution
scales as $\mu \tan \beta M_{\tilde{g}} / M^2_{\tilde{g}, \tilde{b}, \tilde{t}}$ and the SUSY
strongly-interacting sector does not decouple, since the value of the $\mu \tan \beta$ term can be
taken to be much larger than the mass of the SUSY particles appearing at the denominator.
Under these circumstances the study of the Higgs branching fractions, or the Higgs signal strengths,
can unveil SUSY scenarios with strongly-interacting particle at masses well beyond the kinematic reach
of the LHC.

\begin{figure}[hb!]
  \begin{center}
    \vspace*{-0.5cm}
  \begin{tabular}{c}
    \includegraphics[width=0.70\columnwidth]{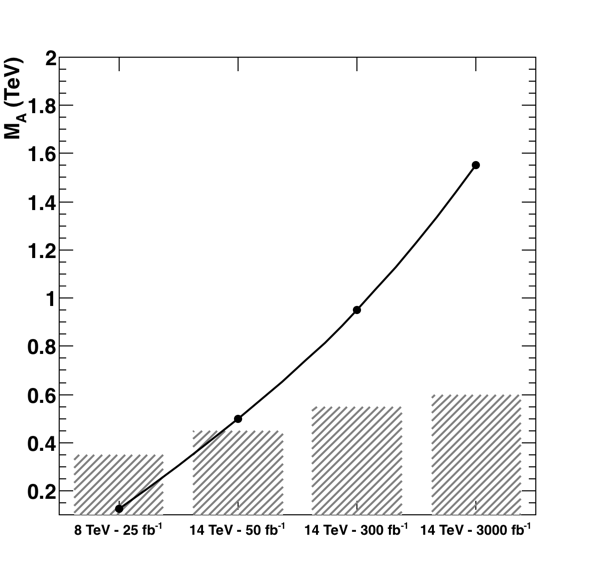} \\
    \vspace*{-0.80cm} \\
  \includegraphics[width=0.70\columnwidth]{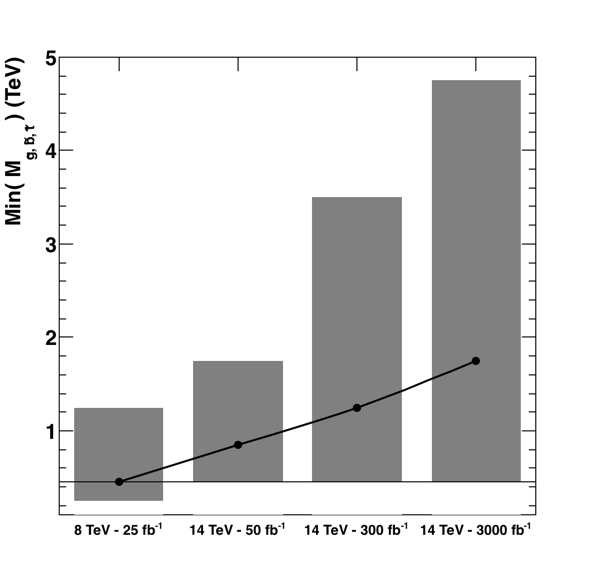} \\
  \end{tabular}
  \vspace*{-1.25cm}
  \end{center}
  \caption{\label{fig:mass-dir-indir-LHC} 
    Sensitivity to the mass of the CP-odd $A$ boson (upper panel) and the lightest
    state among $\tilde{g}$, $\tilde{b}$ and $\tilde{t}$ (lower panel) as a function
    of the energy and luminosity stages of the LHC program for pMSSM points. The 95\%
    C.L.\ exclusion bounds are given for the direct search by the continuous line and
    for the indirect contraints from the $h$ decay properties by the filled bars. These
    show the exclusion range when the MSSM paramaters are varied as described in the text.
    The region shaded in grey has no pMSSM solutions fulfilling the low-energy and flavour
    physics constraints.}
  \vspace*{-0.5cm}
\end{figure}
Results for the sensitivity to SUSY mass scales are summarised in
Figure~\ref{fig:mass-dir-indir-LHC}, which compares the direct and indirect sensitivity
to $M_A$ and to the mass of the gluino and scalar quarks of third generation as a
function of the energy and integrated luminosity of the LHC samples.

In the case of $M_A$, the most stringent constraint is obtained from the analysis of the
Higgs signal strengths and requires $M_A >$ 350~GeV for any value of
$\tan \beta$~\cite{Maiani:2013nga,Arbey:2013jla}. This is due to the weakness of the mass
bound obtained in the $\tau \tau$ channel at low values of $\tan \beta$, where the
product of production cross section and decay branching fraction for
$pp \rightarrow H \rightarrow \tau \tau$ becomes too small to be constrained by the
current LHC data. With the increased energy and luminosity of the Run~2 and then the
HL-LHC program, the direct LHC searches for $A \rightarrow \tau \tau$ but also the
$WW$, $ZZ$ and $tt$ channels, important at small $\tan \beta$ values, are expected to
extend the sensitivity well beyond the value of 600-700~GeV, where $M_A$ effectively
decouples and the indirect sensitivity from the Higgs couplings saturates, independent
of the value of $\tan \beta$. The constraint on $M_A$ is particularly important in
relation to dark matter for predicting the neutralino relic density, since neutralino
annihiliation through the $A$ pole, $\chi \chi \rightarrow A \rightarrow b \bar{b}$,
is a major mechanism for setting the neutralino density in the early universe, when
$M_{\chi} \simeq M_A/2$.

On the contrary, for $\mu \tan \beta$ values of $\cal{O}$(100~TeV), SUSY points with squark
and gluino masses larger than 4~TeV would have the Higgs decay properties deviating enough
from the SM to be identifiable with the accuracy anticipated for the HL-LHC and the ILC, i.e.\
a factor of $\sim$2 larger compared to the anticipated direct sensitivity from MET searches at
the LHC. However, it must be always kept in mind that the indirect sensitivity depends on the
specific values of some parameters, namely $\mu$ and $\tan \beta$, while the direct bounds are
generally much less affected by parametric dependencies. This underlines the importance of the
complementarity of indirect sensitivity through measurements of the Higgs properties
and direct searches~\cite{Cahill-Rowley:2014wba}.

\section{DM and the Higgs sector}

The complementarity between direct searches for SUSY at the LHC, dark matter at direct and
indirect detection experiments has already been highlighted in the framework of the pMSSM
in~\cite{Cahill-Rowley:2014boa}. What has not been discussed in much detail are the relations
between the Higgs sector and dark matter.
If dark matter is due to the lightest SUSY particle (LSP) behaving as a WIMP, Higgs bosons
should in principle couple to it. This has three important consequences. First, the Higgs
can decay into dark matter particle pairs, thus generating a possibly sizeable invisible
Higgs decay width. Then, the Higgs exchange diagram contributes to the scattering cross
section of WIMPs on nucleons and DM particles may annihilate through a Higgs resonance.
Last, there is an important correlation between the WIMP scattering cross section and
the Higgs phenomenology at the LHC and ILC. The parameter space where large SUSY corrections
to the Higgs may be revealed by precision measurements of its decay branching fractions
appears to be complementary to that explorable by the next generation of DM direct
detection experiments. Here we discuss these scenarios in details.

\subsection{Invisible Higgs Decays}

If the lightest LSP neutralino has a mass $M_{\chi} < M_h/2$, then the
$h \rightarrow \chi \chi$ decay channel is kinematically open and the Higgs boson may
acquire a non-zero invisible width.

\begin{figure}[h!]
  \begin{center}
    \vspace*{-0.5cm}
  \begin{tabular}{c}
  \includegraphics[width=0.7\columnwidth]{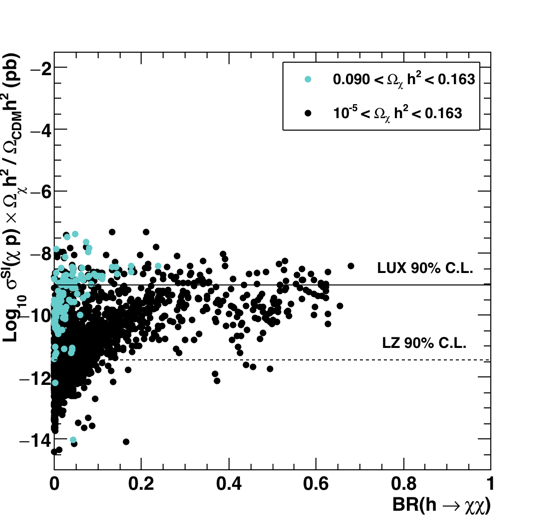} \\
  \end{tabular}
  \vspace*{-1.0cm}
  \end{center}
  \caption{\label{fig:hinvdm}\col 
    Neutralino scattering cross section as a function of its mass for points with
    $\Omega_{\chi} h^2$ compatible with the CMB data (cyan dots) and scaled value for
    those with $\Omega_{\chi} h^2 < \Omega_{CDM} h^2$ (black dots). The horizontal lines
    indicate the current LUX (continuous) and expected LZ (dashed) 90\% C.L. upper limits
    for WIMP masses between 46 and 62~GeV.}
  \vspace*{-0.5cm}
\end{figure}

However, since the $h \chi \chi$ coupling also controls the $\chi p$ scattering cross
section and may contribute to the neutralino annihilation cross section there is a
correlation between the rate of $h \rightarrow \chi \chi$, the scattering cross section
and the neutralino relic density, where the current bounds place some non-trivial
constraints. Figure~\ref{fig:hinvdm} shows this correlation.
SUSY points with neutralino relic density compatible with that extracted from the
analysis of CMB data may have a large Higgs branching fraction into neutralino pair, even
larger than the limit already obtained by the analysis of the Higgs properties at the LHC.
But it is important to observe that a large invisible Higgs branching fraction implies
a large neutralino scattering cross section, because they are both due to an enhanced
$h \chi \chi$ coupling. The LUX upper limit on the cross section for values of the neutralino
mass below $M_h/2$ removes almost entirely the MSSM solutions having BR($h \rightarrow \chi \chi$)
above 0.15, if the neutralino is the only source of dark matter. The LZ experiment will lower
this bound to $\sim$ 0.10 even in the case $\Omega_{\chi} h^2$ does not saturate the relic
density measured on CMB data 
Therefore, if the neutralino LSP alone is responsible for dark matter and the assumptions
on the dark matter density in our galaxy are correct, it is unlikely that the invisible Higgs
rate is very significant, it may even be too small to be directly observable at the LHC. There
has been a significant effort in optimising the LHC sensitivity to invisible Higgs decays, from
fits to rates the observed modes~\cite{Espinosa:2012vu,Belanger:2013kya} to studies of the
associated $VH$ production~\cite{Aad:2014iia,Chatrchyan:2014tja}. However, achieving sensitivity
to invisible rates below 10\% appears very challenging~\cite{Okawa:2013hda}. An $e^+e^-$ collider
will have, on the contrary, the possibility to detect and measure an invisible partial width of
just a few percent~\cite{Schumacher:2003ss}, which makes it unique for exploring this intriguing
scenario.

\subsection{Higgs and DM direct detection}

\begin{figure}[h!]
  \begin{center}
    \vspace*{-0.5cm}
      \begin{tabular}{c}
        \includegraphics[width=0.70\columnwidth]{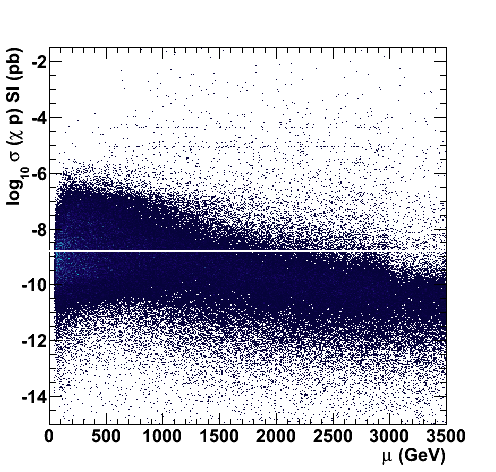} \\
      \end{tabular}
      \vspace*{-1.0cm}
  \end{center}
  \caption{\label{fig:Oh2DDmu}\col 
     Neutralino scattering cross section
     as a function of the Higgsino mass parameter $\mu$ for pMSSM points. The horizontal line
     represents the current LUX bound.}
  \vspace*{-0.5cm}
\end{figure}

\begin{figure}[h!]
  \vspace*{-0.5cm}
  \begin{center}
    \includegraphics[width=0.70\columnwidth]{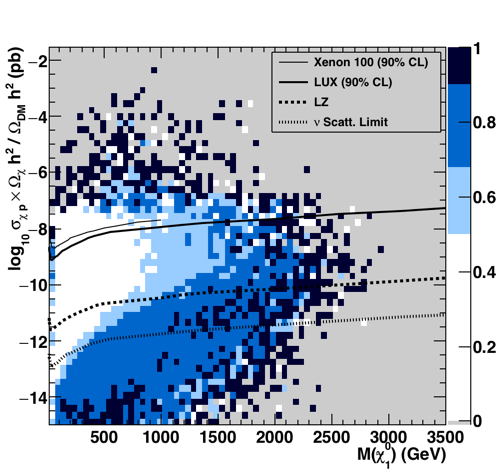} \\
  \end{center}
  \vspace*{-1.0cm}
  \caption{\label{fig:DDN1hBR}\col 
    Fraction of pMSSM point not excluded by the LHC 7+8~TeV data and excluded at 95\% C.L. by the
    analysis of the Higgs branching fractions with the accuracy expected at a 1~TeV ILC in the plane defined
    by the neutralino mass and the scaled neutralino scattering cross section compared to the scattering
    cross section current and future upper limits. The colours show the regions
    where more than 50\% (light blue), 68\% (blue) and 90\% (dark blue) of the points are excluded.}
  \vspace*{-0.5cm}
\end{figure}

If the neutralino WIMP is heavier than half the Higgs mass, the invisible $h \rightarrow \chi \chi$
decay is forbidden. Still, the Higgs coupling to the neutralino is relevant to dark matter
phenomenology. The WIMP scattering cross section receives contributions from the Higgs exchange
and is inversely proportional to the Higgsino mass parameter, $\mu$, as illustrated in
Figure~\ref{fig:Oh2DDmu}.
Therefore, there are two complementary regions in the SUSY parameter space: the first, where
$\mu \tan \beta$ is large, the neutralino scattering cross section is small, the strongly-interacting
SUSY particles do not decouple and the Higgs branching fractions remain significantly different
from their SM predictions, and the second, where the neutralino scattering cross section is large
enough to be tested by dark matter direct detection experiments while the heavy SUSY degrees of
freedom decouple in the Higgs sector and the Higgs properties are SM-like. This has the important
consequence that, in case neither the LHC and, possibly, the ILC studies of the Higgs profile would
reveal a deviation from the SM, nor the LZ data would detect a WIMP signal, a very significant
fraction of the parameter space of the MSSM with neutralino LSP could be excluded
(see Figure~\ref{fig:DDN1hBR}). 

This is demostrated in quantitative terms in Table~\ref{tab:methiggs}, which summarises the fraction of
our pMSSM points with SUSY masses up to 5~TeV excluded by the LHC MET searches only and by the addition of
the Higgs data (at LHC and ILC) and the DM direct searches for the loose $\Omega h^2$ constraints. Requiring
the neutralino to saturate the relic density decreases the fraction of points excluded by the LHC at 14~TeV
by about 12\% due to the shift of wino $\chi^0_1$ points towards larger masses, but does not significantly affect
the overall results when the Higgs and direct dark matter constraints are applied. Independent of the LHC energy
and statistics considered, the inclusion of the Higgs data always significantly increases the fraction of SUSY
scenarios which can be tested and excluded, if the Higgs properties turn out to be SM-like. The combined analysis
of the LHC, ILC and LZ data, assuming no signal is observed, should provide bounds stringent enough to
exclude almost 97\% of the pMSSM points we have generated and having SUSY masses up to 5~TeV. If a signal
is observed, crucial tests can be carried to ensure that the nature of the WIMP dark matter is identified.

\begin{table}
  \begin{andptabular}[\columnwidth]{|c|c|c|c|c|}{\label{tab:methiggs} Fractions of pMSSM points
      excluded by the combination of LHC MET searches, LHC and ILC Higgs
      data and LZ DM direct detection}%
    & LHC 8  & LHC & LHC & HL-LHC \\
    & 8~TeV  & 14~TeV & 14~TeV & 14~TeV \\
    & 25 fb$^{-1}$   &  50~fb$^{-1}$ & 300 fb$^{-1}$    & 3 ab$^{-1}$ \\
    js+$\ell$s+MET   & 0.145      &  0.570 & 0.698      & 0.820  \\
    +h$^0$ $\mu$s    & 0.317      &  0.622 & 0.793      & 0.920 \\
    +ILC h$^0$ BRs   & 0.588      &  0.830 & 0.890      & 0.945 \\
    +LZ              &            &  0.914 & 0.940      & 0.964 \\    
  
  \end{andptabular}
  \vspace*{-0.5cm}
\end{table}
The complementarity between the indirect probe through the Higgs measurements and the direct
SUSY searches in the MET channels at colliders is thus integrated by the DM direct detection data.
The anticipated precision of an $e^+e^-$ collider is crucial in this context. Together, all these
data could test in details supersymmetry as the theory of new physics in relation to dark matter.

\section{Conclusions}

The discovery of a light Higgs boson at the LHC constitutes a major step towards the experimental
test of models of new physics and the investigation of their relation to dark matter. If this is due
to a WIMP, then the Higgs boson most likely couples to it and to the other particles of the underlying
SM extension. Supersymmetry represents a well motivated implementation of
new physics incorporating a light Higgs boson and a dark matter candidate.

Dark matter introduces a number of important constraints on the SUSY parameter space when the neutralino
is the WIMP and bound by dark matter data. The contribution of the Higgs to the neutralino scattering cross
section offers opportunities to test SUSY which are complementary to those available at colliders.

SUSY contributions which shift the coupling of the Higgs to $b$ quarks and therefore modify all its
decay branching fractions do not decouple for large values of the mass of the strongly interacting
particles if the product of $\mu \tan \beta$ is large. Therefore, Higgs precision measurements retain
their sensitivity to SUSY corrections, even if the direct searches in the MET channels with the 14~TeV
LHC data do not obtain signals of SUSY states.

The complementarity between the indirect probe through precision Higgs measurements, the direct
SUSY searches in the MET channels at colliders is also integrated by the DM direct detection data.
This complementarity is highlighted by the observation that the region of parameter space where the
Higgs couplings and the WIMP neutralino scattering can give signals beyond the SM are largely
complementary. The anticipated precision of an $e^+e^-$ collider is crucial to maximise this
complementarity.  There are excellent perspectives that the combination of these data will test
in details supersymmetry as the theory of new physics, if this is indeed reponsible for dark matter.


\begin{thebibliography}{99}

\bibitem{ATLAS:2012zz}
  G.~Aad {\it et al.}  [ATLAS Collaboration],
  Phys.\ Lett.\ B {\bf 716} (2012) 1
  [arXiv:1207.7214 [hep-ex]].
  
\bibitem{CMS:2012zz}
  S.~Chatrchyan {\it et al.}  [CMS Collaboration],
  Phys.\ Lett.\ B {\bf 716} (2012) 30
  [arXiv:1207.7235 [hep-ex]].

\bibitem{Aprile:2012nq}
  E.~Aprile {\it et al.}  [XENON100 Collaboration],
  Phys.\ Rev.\ Lett.\  {\bf 109} (2012) 181301
  [arXiv:1207.5988 [astro-ph.CO]].
  
\bibitem{Akerib:2013tjd}
  D.~S.~Akerib {\it et al.}  [LUX Collaboration],
  Phys.\ Rev.\ Lett.\  {\bf 112} (2014) 9,  091303
  [arXiv:1310.8214 [astro-ph.CO]].

\bibitem{Dawson:2013bba}
  S.~Dawson {\it et al.},
  arXiv:1310.8361 [hep-ex].
  
\bibitem{Malling:2011va}
  D.~C.~Malling {\it et al.},
  arXiv:1110.0103 [astro-ph.IM].

\bibitem{Cushman:2013zza}
  P.~Cushman {\it et al.},
  arXiv:1310.8327 [hep-ex].

\bibitem{Barletta:2014vea}
  W.~Barletta, M.~Battaglia, M.~Klute, M.~Mangano, S.~Prestemon, L.~Rossi and P.~Skands,
  Nucl.\ Instrum.\ Meth.\ A {\bf 764} (2014) 352.

\bibitem{Barletta:2014nka}%
  W.~A.~Barletta {\it et al.},
  arXiv:1401.6114 [hep-ex].

\bibitem{Behnke:2013xla}
  T.~Behnke {\it et al.},
  arXiv:1306.6327 [physics.acc-ph].

\bibitem{Zimmermann:2014qxa}
  F.~Zimmermann, M.~Benedikt, D.~Schulte and J.~Wenninger,
  IPAC-2014-MOXAA01.
  
\bibitem{Djouadi:1998di}
  A.~Djouadi {\it et al.},  hep-ph/9901246.
  
\bibitem{AbdusSalam:2009qd}
  S.~S.~AbdusSalam {\it et al.}
  Phys.\ Rev.\ D {\bf 81} (2010) 095012
  [arXiv:0904.2548 [hep-ph]].

\bibitem{Sekmen:2011cz}
  S.~Sekmen {\it et al.},
  JHEP {\bf 1202} (2012) 075
  [arXiv:1109.5119 [hep-ph]].

\bibitem{Arbey:2011un}
  A.~Arbey, M.~Battaglia and F.~Mahmoudi,
  Eur.\ Phys.\ J.\ C {\bf 72} (2012) 1847
  [arXiv:1110.3726 [hep-ph]].

\bibitem{Arbey:2012dq}
  A.~Arbey, M.~Battaglia, A.~Djouadi and F.~Mahmoudi,
  JHEP {\bf 1209} (2012) 107
  [arXiv:1207.1348 [hep-ph]].

\bibitem{CahillRowley:2012kx}
  M.~W.~Cahill-Rowley, J.~L.~Hewett, A.~Ismail and T.~G.~Rizzo,
  Phys.\ Rev.\ D {\bf 88} (2013) 3,  035002
  [arXiv:1211.1981 [hep-ph]].

\bibitem{Arbey:2012bp}
  A.~Arbey, M.~Battaglia, A.~Djouadi and F.~Mahmoudi,
  Phys.\ Lett.\ B {\bf 720} (2013) 153
  [arXiv:1211.4004 [hep-ph]].

\bibitem{Cahill-Rowley:2014boa}
  M.~Cahill-Rowley {\it et al.},
  Phys.\ Rev.\ D {\bf 91} (2015) 5,  055011
  [arXiv:1405.6716 [hep-ph]].
  
\bibitem{CMS:2013rda}
  CMS Collaboration,
  CMS-PAS-SUS-12-030.
  
\bibitem{CMS:2014mia}
  CMS Collaboration,
  CMS-PAS-SUS-13-020.
  





  




\bibitem{Ade:2013zuv}
  P.~A.~R.~Ade {\it et al.}  [Planck Collaboration],
  Astron.\ Astrophys.\  (2014)
  [arXiv:1303.5076 [astro-ph.CO]].
  
\bibitem{ATLAS-CONF-2013-047}
  [ATLAS Collaboration], Note ATLAS-CONF-2013-047.

\bibitem{ATLAS-CONF-2013-053}
  [ATLAS Collaboration], Note ATLAS-CONF-2013-053.

\bibitem{Aad:2014qaa}
  G.~Aad {\it et al.}  [ATLAS Collaboration],
  JHEP {\bf 1406} (2014) 124
  [arXiv:1403.4853 [hep-ex]].

\bibitem{ATLAS-CONF-2013-037}
  [ATLAS Collaboration], Note ATLAS-CONF-2013-037.

\bibitem{ATLAS-CONF-2013-049}
  [ATLAS Collaboration], Note ATLAS-CONF-2013-049.

\bibitem{ATLAS-CONF-2013-035}
  [ATLAS Collaboration], Note ATLAS-CONF-2013-035.

\bibitem{ATLAS-CONF-2013-093}
  [ATLAS Collaboration], Note ATLAS-CONF-2013-093.

\bibitem{CMS-EXO-12-048}
  [CMS Collaboration], Note CMS PAS EXO-12-048.
    
\bibitem{ATLAS-CONF-2013-068}
  [ATLAS Collaboration], Note ATLAS-CONF-2013-068.

\bibitem{CMS-HIG-13-021}
  [CMS Collaboration], Note CMS PAS HIG-13-021.
  
\bibitem{Read:2002hq}
  A.~L.~Read,
  J.\ Phys.\ G {\bf 28} (2002) 2693.

\bibitem{Djouadi:2005gj}
  A.~Djouadi,
  Phys.\ Rept.\  {\bf 459} (2008) 1
  [hep-ph/0503173].

\bibitem{Muhlleitner:2010zz}
  M.~Muhlleitner, H.~Rzehak and M.~Spira,
  DESY-PROC-2010-01.

\bibitem{Maiani:2013nga}
  L.~Maiani, A.~D.~Polosa and V.~Riquer,
  Phys.\ Lett.\ B {\bf 724} (2013) 274
  [arXiv:1305.2172 [hep-ph]].

\bibitem{Arbey:2013jla}
  A.~Arbey, M.~Battaglia and F.~Mahmoudi,
  Phys.\ Rev.\ D {\bf 88} (2013) 1,  015007
  [arXiv:1303.7450 [hep-ph]].
  
\bibitem{Cahill-Rowley:2014wba}
  M.~Cahill-Rowley, J.~Hewett, A.~Ismail and T.~Rizzo,
  Phys.\ Rev.\ D {\bf 90} (2014) 9,  095017
  [arXiv:1407.7021 [hep-ph]].

\bibitem{Espinosa:2012vu}
  J.~R.~Espinosa, M.~Muhlleitner, C.~Grojean and M.~Trott,
  JHEP {\bf 1209} (2012) 126
  [arXiv:1205.6790 [hep-ph]].

\bibitem{Belanger:2013kya}
  G.~Belanger {\it et al.},
  Phys.\ Lett.\ B {\bf 723} (2013) 340
  [arXiv:1302.5694 [hep-ph]].

\bibitem{Aad:2014iia}
  G.~Aad {\it et al.}  [ATLAS Collaboration],
  Phys.\ Rev.\ Lett.\  {\bf 112} (2014) 201802
  [arXiv:1402.3244 [hep-ex]].

\bibitem{Chatrchyan:2014tja}
  S.~Chatrchyan {\it et al.}  [CMS Collaboration],
  Eur.\ Phys.\ J.\ C {\bf 74} (2014) 8,  2980
  [arXiv:1404.1344 [hep-ex]].

\bibitem{Okawa:2013hda}
  H.~Okawa, J.~Kunkle and E.~Lipeles,
  arXiv:1309.7925 [hep-ex].

\bibitem{Schumacher:2003ss}
  M.~Schumacher,
  LC-PHSM-2003-096.
  
\end{thebibliography}
\end{document}